\documentstyle[titlepage,12pt]{article}
\long\def\@makefntext#1{\parindent 0cm\noindent
\hbox to 1em{\hss$^{\@thefnmark}$}#1}
\def\n{\nonumber\\}
\def\6j#1#2{\left\{ \begin{array}{ccc} #1 \\ #2 \end{array} \right\}}
\def\q6j#1#2{\left\{ \begin{array}{ccc} #1 \\ #2 \end{array}
\right\}_{\textstyle \! q}}
\def\b#1{[#1]_q}
\def\l{\lambda}
\def\12{\frac{1}{2}}
\def\fq#1{F_q(#1)}
\def\fc#1{F_c(#1)}
\def\qc{\epsilon_q}
\begin{document}
%
%
\input{psfig}
\begin{titlepage}
\hspace*{12cm}%
YITP/U- 91-43

\hspace*{12cm}%
October, 1991

\vspace*{2cm}

\addtocounter{footnote}{1}

\begin{center}

{\large \bf 3-dimensional Gravity from the Turaev-Viro Invariant}
\\
\vspace{0.5cm}

\vspace{1cm}
{\sc Shun'ya Mizoguchi}\footnote{JSPS fellow.
e-mail: mizo@jpnrifp.yukawa.kyoto-u.ac.jp}

\vspace{0.5cm}
and

\vspace{0.5cm}

{\sc Tsukasa Tada}\footnote{Soryuushi shogakukai fellow.
e-mail: tada@yisun1.yukawa.kyoto-u.ac.jp}

\vspace{1cm}
{\it Uji Research Center \\
Yukawa Institute for Theoretical Physics \\
Kyoto University, Uji 611 Japan \\}

\vspace{1cm}
\vspace{2cm}

{\bf ABSTRACT}
\end{center}
We study the $q$-deformed su(2)
spin network as a 3-dimensional quantum gravity model.
We show that in the semiclassical continuum limit
the Turaev-Viro invariant obtained recently defines
naturally regularized path-integral $\grave{\rm a}$ la
Ponzano-Regge, In which a contribution from the cosmological
term is effectively included. The regularization dependent
cosmological constant is found to be ${4\pi^2\over k^2}
+O(k^{-4})$, where $q^{2k}=1$. We also discuss the relation to
the Euclidean Chern-Simons-Witten gravity in 3-dimension.

\end{titlepage}
\baselineskip = 0.7cm

	It is well-known that 3-dimensional gravity is perturbatively
trivial in the sense that there are no local degrees of freedom.
In 3-dimension Ricci-flat space-time means truly flat space-time,
so that there are no gravitational wave modes but is only topological
excitation. Topological nature of 3-dimensional gravity has already
been encountered in the old work by Ponzano and Regge
on the semi-classical limit of the Racah coefficients of su(2)\cite{PR}.
In their seminal paper (and also ref.\cite{HP})
a triangulation-independent quantity was defined
by utilizing a certain relation for $6j$ symbols, and was shown to be
viewed as a path-integral for 3-dimensional
quantum gravity in the semi-classical continuum limit,
though their expression diverges and needs some regularization.
Recently, Turaev and Viro has constructed a new topological invariant
from the $q$-deformed su(2) spin network when $q$ is a root of unity.
Their construction strongly resembles that of ref.\cite{PR},
and moreover, the Turaev-Viro (TV) invariant is naturally
regularized and finite due to the restriction for the spin variables.
Therefore the TV invariant is expected to be considered
as a regularized path-integral for 3-dimensional quantum gravity.

	In this letter we estimate the asymptotic behavior
of the $q$-$6j$ symbol by the WKB approximation along
the method of ref.\cite{SG},
and see how the path-integral defined by the su(2) spin network
receives `quantum' corrections from $q$-deformation.
We show that a contribution from the cosmological term is effectively
included in the path-integral from the TV invariant.

	In ref.\cite{PR} a sum of products of four (classical)
su(2) Racah-Wigner $6j$ symbols was considered:
\begin{eqnarray}
&\displaystyle{\sum_{\scriptsize
\begin{array}{c} x,l_1,l_2,l_3:\\ \mbox{\rm allowed value}\end{array}
}}&
(2x+1)(2l_1+1)(2l_2+1)(2l_3+1)~(-1)^{\chi}\nonumber \\
&&\hskip -10mm
\times\left\{
\begin{array}{ccc}
j_1&j_2&j_3 \\ x&l_1&l_2
\end{array}
\right\}
\left\{
\begin{array}{ccc}
j_6&j_5&j_1 \\ l_1&l_2&l_3
\end{array}
\right\}
\left\{
\begin{array}{ccc}
j_4&j_2&j_6 \\ l_2&l_3&x
\end{array}
\right\}
\left\{
\begin{array}{ccc}
j_3&j_5&j_4 \\ l_3&x&l_1
\end{array}
\right\}, \label{eq:four6j}
\end{eqnarray}
where $\chi=x+\sum^3_{i=1}l_i+\sum^6_{i=1}j_i$.
By repeated application of the Biedenharn-Elliott identity
and the orthogonal relation,
(\ref{eq:four6j}) is reduced to a single $6j$ symbol:
\begin{equation}
\sum_{x=0}^{\infty}(2x+1)^2
\left\{
\begin{array}{ccc}
j_1&j_2&j_3 \\ j_4&j_5&j_6
\end{array}
\right\}.
\label{eq:subdivision}
\end{equation}
To give a meaning to the divergent expression (\ref{eq:subdivision})
a large angular-momentum cut-off $R$
is introduced in the summation, so that we take
the equality between (\ref{eq:four6j}) and (\ref{eq:subdivision})
as the following renormalized identity:
\begin{eqnarray}
&&\left\{
\begin{array}{ccc}
j_1&j_2&j_3 \\ j_4&j_5&j_6
\end{array}
\right\}\nonumber\\
&=&\lim_{R\rightarrow\infty}
\frac{3a}{4R^3}
\sum_{
x,l_1,l_2,l_3<R
}
(2x+1)(2l_1+1)(2l_2+1)(2l_3+1)~(-1)^{\chi}\nonumber \\
&&\times\left\{
\begin{array}{ccc}
j_1&j_2&j_3 \\ x&l_1&l_2
\end{array}
\right\}
\left\{
\begin{array}{ccc}
j_6&j_5&j_1 \\ l_1&l_2&l_3
\end{array}
\right\}
\left\{
\begin{array}{ccc}
j_4&j_2&j_6 \\ l_2&l_3&x
\end{array}
\right\}
\left\{
\begin{array}{ccc}
j_3&j_5&j_4 \\ l_3&x&l_1
\end{array}
\right\}, \label{eq:rensubdivision}
\end{eqnarray}
where $a$ is some dimensionless constant (finite renormalization).
Since a $6j$ symbol is naturally associated to a tetrahedron,
the identity (\ref{eq:rensubdivision})
means that the value of $6j$ associated to a tetrahedron is invariant
under subdivision if the summation is carried out over all allowed
lengths of internal edges with appropriate weight (Fig.1).
Repeating the subdivision (\ref{eq:rensubdivision}), we find
\begin{eqnarray}
&&\left\{
\begin{array}{ccc}
j_1&j_2&j_3 \\ j_4&j_5&j_6
\end{array}
\right\}\n
&=&\lim_{R\rightarrow\infty}
(\frac{3a}{4R^3})^e
\sum_{x_1<R}\cdots\sum_{x_f<R}\prod_{i=1}^d[T_i]\nonumber\\
&&\hskip 10mm
\times (-1)^{\sum_{l=1}^6 j_l}\prod_{p=1}^f (-1)^{x_p}(2x_p+1) \n
&\equiv&I_{PR}~,
\end{eqnarray}
where
\begin{eqnarray}
[T_i]&:&\mbox{\rm $6j$ symbol assigned to the $i$th simplex} \n
d,\ e,\ f&:&\mbox{\rm the number of simplices,internal vertices } \n
&&\mbox{and internal edges,respectively } \n
x_p&:&(\mbox{\rm the length of the $p$th edge})-\frac{1}{2}. \nonumber
\end{eqnarray}
Hence $I_{PR}$ may be taken as a `topological invariant',
where topological invariant
means that it does not depend on the triangulation inside
the 3-manifold,
but depends only on the edge-lengths on the boundary and topology
of the triangulated 3-manifold. Obviously, with fixed boundaries,
$I_{PR}$ takes the same value in each class of triangulation
connected through the operation (\ref{eq:rensubdivision}).

	In the semi-classical continuum limit
$x_p$, $e$ (and so $f$)
go to infinity. The $6j$ symbol behaves as \cite{PR,SG}
\begin{equation}
\left\{
\begin{array}{ccc}
j_1&j_2&j_3 \\ j_4&j_5&j_6
\end{array}
\right\}
\approx
(\frac{1}{12\pi V})^{\frac{1}{2}}
\cos(\sum_{i=1}^f\theta_iJ_i+\frac{\pi}{4})
\end{equation}
in the domain where $J_i$ are uniformly large (see Figure Caption).
Here $V$ is the volume of the tetrahedron.
Thus one may write $I_{PR}$ as
\begin{equation}
I_{PR}\sim\int d\mu \prod_{\mbox{\scriptsize\rm tetrahedra}}
\cos\left({\sum_{i=1}^f x_i(\pi-\theta_i)}-\frac{\pi}{4}\right)
\end{equation}
with
\begin{equation}
d\mu=(\frac{3a}{4R^3})^e\prod_{i=1}^fdx_i (2x_i+1).
\end{equation}
Taking only the positive frequency part of cosine, together with
$\sum_{i=1}^f x_i(\pi-\theta_i)\sim\frac{1}{2}\int\sqrt{g}R$
\cite{Regge}, one finds
\begin{equation}
I_{PR}~(\mbox{\rm positive frequency part})
\sim \int d\mu \exp\left(i\frac{1}{2}\int\sqrt{g}R\right)~.
\label{eq:pathintegral}
\end{equation}
The above consideration is an interesting possibility
to relate $I_{PR}$ with 3d quantum
gravity, though there are some difficulties, {\it i.e.}
the appearance of $i$ in the exponent and the treatment of the
other interference terms.

	Turaev and Viro defined the quantity $I_{TV}$
(up to the factors from the boundary) as \cite{TV}
\begin{eqnarray}
I_{TV}&\equiv&w^{-2e}\sum_{\scriptsize\mbox{\rm allowed $\!\phi$}}
\prod_{i=1}^d |T_i^{\phi}|
\prod_{i=1}^f w_i^2~,
\end{eqnarray}
where $|T_i^{\phi}|$, $w_i$ and $w$ are weight assigned
to tetrahedra, edges and vertices, respectively.
$\phi$ stands for a configuration of edge-lengths,
called {\em coloring}.
$\phi$ is an allowed (`admissible' in ref.\cite{TV}) coloring
if any set of triple which forms a face of a tetrahedron
satisfies the triangle inequality,
and if the sum of any triple is an integer less than $k-2$.
They proved that $I_{TV}$ is independent of triangulation if
\begin{eqnarray}
&&|T_i^{\phi}|=(-1)^{\sum_{l=1}^6 j^{(i)}_l}
\left\{
\begin{array}{ccc}
j^{(i)}_1&j^{(i)}_2&j^{(i)}_3 \\ j^{(i)}_4&j^{(i)}_5&j^{(i)}_6
\end{array}
\right\}_{\textstyle q} \n
&&w_i^2=[2x_i+1]_q,\hskip 5mm w^2=-\frac{2k}{(q-q^{-1})^2},
\end{eqnarray}
where 
$\bigl\{ \mathop{}^{\cdots}_{\cdots} \bigr\}$ stands for a (restricted)
$q$-$6j$ symbol with $q=e^{\frac{\pi i}{k}}$
\cite{KR}, and $[n]_q$ is a $q$-integer
$[n]_q=\frac{q^n-q^{-n}}{q-q^{-1}}$.
If $q$ is a root of unity in general,
 some Clebsch-Gordan coefficients of the representation
of $U_q$(sl(2)) diverge. Therefore one must,
and can successfully, restrict the domain of the spin variables
to a finite set ($\{0,\frac{1}{2},1,\ldots,k-\frac{1}{2}\}$
in our case).

It is surprising that $I_{PR}$ and $I_{TV}$ are in the same form.
Moreover, since
\begin{eqnarray}
\left\{
\begin{array}{ccc}
j^{(i)}_1&j^{(i)}_2&j^{(i)}_3 \\ j^{(i)}_4&j^{(i)}_5&j^{(i)}_6
\end{array}
\right\}_{\textstyle q}
&=&\left\{
\begin{array}{ccc}
j^{(i)}_1&j^{(i)}_2&j^{(i)}_3 \\ j^{(i)}_4&j^{(i)}_5&j^{(i)}_6
\end{array}
\right\}+O(k^{-2}) \n
{[2x_i+1]}_q&=&2x_i+1+O(k^{-2}) \n
-\frac{2k}{(q-q^{-1})^{2}}&=&\frac{k^3}{2\pi^2}(1+O(k^{-2}))~,
\end{eqnarray}
$I_{TV}$ approaches $I_{PR}$ in the $k\rightarrow\infty$
($q\rightarrow 1$) limit with $a=\frac{8\pi^2}{3}$,
where $k$ plays the same role as $R$.
In other word, the TV invariant $I_{TV}$ provides a natural
regularized path-integral for 3-dimensional quantum gravity.

	Now let us study the asymptotic behavior
of $q$-$6j$ symbols. We start from the following relation,
which is a special case of quantum analogue of
the Biedenharn-Elliot identity for restricted $q$-$6j$ symbols
\cite{KR}\footnote{
Nomura also discussed identities of $q$-$6j$
and asymptotic behavior when a part of the arguments is  large
\cite{Nomura1}.}:

\begin{eqnarray}
 && \q6j{j_1&j_2&j_3}{j_4&j_5&j_6} \q6j{j_1&j_2&j_3}{\12&j_3+\Delta_b&
 j_2+\Delta_a}\nonumber \\
 && = \sum_{\zeta=j_4 \pm \12}
 (-1)^{j_1+2 j_2+2 j_3+j_4+j_5+j_6+\zeta +\12 +\Delta_a+\Delta_b}
 \b{2 \zeta +1}\nonumber \\
  && \times
 \q6j{j_1&j_2+\Delta_a&j_3+\Delta_b}{\zeta&j_5&j_6}
 \q6j{\12&j_2&j_2+\Delta_a}
 {j_6&\zeta&j_4}\nonumber\\&&\qquad \times
\q6j{\12&j_3+\Delta_b&j_3}{j_5&j_4&\zeta}
 \qquad \Bigl( \Delta_{a,b}=\pm \12 \Bigr) .
 \label{eqn:qBiHa}
\end{eqnarray}
Combining the equations above,
one obtains the following recursion relation:
\begin{eqnarray}
 & &\b{2j_1}g(j_1+1)\q6j{j_1+1&j_2&j_3}{j_4&j_5&j_6}
 \nonumber \\ & &+2h(j_1)
 \q6j{j_1&j_2&j_3}{j_4&j_5&j_6} \nonumber \\
 & &
 +\b{2j_1+2}g(j_1)\q6j{j_1-1&j_2&j_3}{j_4&j_5&j_6}
 =0 \label{eqn:qBiEl}\\
 & &\hfill \nonumber
 \\
 &&g(j)=\bigl\{\b{j_2+j_3+1+j}\b{j_2+j_3+1-j}
 \nonumber \\& & \qquad \quad \; \times
 \b{j+j_2-j_3}\b{j-j_2+j_3} \nonumber\\
 &&\qquad \quad \; \times \b{j_5+j_6+1+j}\b{j_5+j_6+1-j}
 \nonumber \\ & &\qquad \quad \; \times \b{j+j_5-j_6}
 \b{j-j_5+j_6}\bigr\}^{1\over2} \qquad \\
 &&h(j)=\Bigl\{\b{2j+1}\b{2j+2}\b{2j}\nonumber \\
 &&\qquad \qquad \qquad \times\b{j_3+j_4+j_5+2}
 \b{j_3+j_5-j_4+1} \nonumber \\
 & &\qquad  -\b{2j}\b{j+j_5+j_6+2}\b{j+j_5-j_6+1}\nonumber
 \\ && \qquad \qquad \qquad \times \b{j+j_2+j_3+2}
 \b{j+j_3-j_2+1} \nonumber \\
 &&\qquad -\b{2j+2}\b{j_6-j_5+j}\b{j_5+j_6-j+1}
 \nonumber \\ && \qquad \qquad \qquad \times \b{j_2+j_3-j+1}
 \b{j-j_2+j_3} \Bigr\}/2~.
\end{eqnarray}

Since we are considering restricted $q$-$6j$ symbols,
the spin variables are truncated
at $k$. Therefore, to take large angular
momentum limit we have to take $k$
also large. Hence we substitute
$j_1,j_2,\ldots$ for $\l j_1, \l j_2,\ldots$
and $k$ for $\l k$ at the same time,
then we send \(\l\rightarrow\infty\),
keeping the ratio $j_i/k \ll 1$.
Defining the $q$-analog of triangle-area:
\begin{equation}
 \fq{a,b,c}=\frac 14 \sqrt{\b{a+b+c}\b{a-b+c}\b{a+b-c}\b{-a+b+c}}~,
 \label{eqn:qheron}
\end{equation}
we can express \(g(j)\) in a more compact form:
\begin{equation}
 g(j)=16\fq{j,j_2+\12,j_3+\12}\fq{j,j_5+\12,j_6+\12}.
\end{equation}
Recalling the distinction between $J$ and $j$ (Fig.1),
we obtain
\begin{eqnarray}
 g(\l j)&=&16\sqrt{\fq{\l J,\l J_2,\l J_3}\fq{\l J \! -\!
 1,\l J_2,\l J_3} \fq{\l J,\l J_5,\l J_6}\fq{\l J\! - \! 1,
 \l J_5,\l J_6}}  \nonumber \\
 &&\quad \times \Bigl[ 1+ O({\l^{-2}})\Bigr].
\end{eqnarray}
Thus we  rewrite the relation (\ref{eqn:qBiEl}), keeping
terms up to order $\l^{-1}$, as follows:
\begin{eqnarray}
 & &\left\{ \frac{\fq{\l J_1+1,\l J_2,\l J_3}
 \fq{\l J_1+1,\l J_5,\l J_6}}
 {\b{2(\l J_1 +1)}}\right\}^{\12}
 \q6j{\l j_1 +1&\l j_2&\l j_3}{\l j_4&\l j_5&\l j_6} \nonumber \\
 &+&\left\{ \frac{\fq{\l J_1-1,\l J_2,\l J_3}
 \fq{\l J_1-1,\l J_5,\l J_6}}{\b{2(\l J_1 -1)}}\right\}^{\12}
 \q6j{\l j_1 -1&\l j_2&\l j_3}{\l j_4&\l j_5&\l j_6}  \nonumber \\
 &+& \frac{h(\l j_1)}{8\left\{ \b{2(\l J_1 \! +\! 1)}\b{2
 (\l J_1\! -\! 1)} \right\}^{1/2}}
 \frac{\left\{\fq{\l J_1,\l J_2,\l J_3}
 \fq{\l J_1,\l J_5,\l J_6}\right\}^{-1/2}}
 {\left\{\b{2 \l J_1}\right\}^{1/2}}
 \nonumber \\
& & \qquad \times \q6j{\l j_1&\l j_2&\l j_3}
 {\l j_4&\l j_5&\l j_6}
 =0. \label{eqn:mqBiEl}
\end{eqnarray}
Introducing
\begin{eqnarray}
 \varphi(j_1) &\equiv&
 \left\{ \frac{\fq{\l J_1,\l J_2,\l
 J_3}\fq{\l J_1,\l J_5,\l J_6}}
 {\b{2 \l J_1}}\right\}^{1/2}
 \q6j{\l j_1&\l j_2&\l j_3}{\l j_4&\l j_5&\l j_6}
 \label{eqn:phidef} \\
 c(J_1) &\equiv&
 -\frac{h(\l j_1)}{16\left\{ \b{2(\l J_1+1)}
 \b{2(\l J_1-1)}\right\}^{1/2}
 F_q(\l J_1, \l J_2, \l J_3)F_q(\l J_1, \l J_5, \l J_6)}~,
 \label{eqn:cdef}
\end{eqnarray}
we arrive at a difference equation for \(j_1\):
\begin{equation}
 \left\{\Delta^2 + 2 -2c(J_1)\right\} \varphi(j_1)=0.
 \label{eqn:difeq}
\end{equation}

	Let us estimate the solution of the above equation
(\ref{eqn:difeq}) in the large \(k\) limit.
When \(q\rightarrow1\), \(c(J_1)\) approaches its `classical' value
\(\cos \theta_1\) \cite{SG},
so one expands \(c(J_1)\) as
\begin{equation}
c(J_1)=\cos\theta_1+\varrho+O(k^{-4})~,
\end{equation}
where the next-leading term $\varrho$
is of order \({k^{-2}}\). Since
\begin{eqnarray}
 \cos(\theta -\frac{\varrho}{\sin \theta}) &=&
 \cos \theta \cos (\frac{\varrho}{\sin \theta})
 +\sin \theta \sin (\frac{\varrho}{\sin \theta}) \nonumber\\
 &=& \cos \theta \left( 1-\frac{\varrho^2}{2 \sin^2 \theta} +
 \ldots \right) + \sin \theta \left( \frac{\varrho}{\sin \theta}
 -\frac{\varrho^3}{3! \sin^3 \theta}+\ldots \right) \nonumber\\
 &=& \cos \theta + \varrho + O(k^{-4})~,
\end{eqnarray}
(\ref{eqn:difeq}) becomes in the large $k$ limit as follows:
\begin{equation}
 \left\{\Delta^2 + 2
 -2\cos (\theta _1 -\frac{\varrho}{\sin \theta_1})
 \right\} \varphi(x)=0.
 \label{eqn:difeqk2}
\end{equation}
Here we have changed the variable from \( J_1\) to \(x\).
According to ref.\cite{SG}, we solve the difference equation
(\ref{eqn:difeqk2}) by WKB approximation. The result is
\begin{equation}
 \varphi (x) \approx \frac{C}{\sqrt{\sin (\theta _1 -
 \frac{\varrho}{\sin \theta_1})}} \cos \left( \int
 (\theta _1 -\frac{\varrho}{\sin \theta_1}) dx +
 \frac{\pi}{4} \right),
\end{equation}
where \(C\) is a normalization.
Thus  we obtain the expression for the asymptotic formula of $q$-$6j$
symbol:
\begin{equation}
 \q6j{j_1&j_2&j_3}{j_4&j_5&j_6} \sim \frac{C'}{\sqrt V}
 \cos \left( \int \theta_1 (x) dx -
 \int \frac{\varrho (x)}{\sin \theta_1} dx +\frac{\pi}{4} \right).
\label{eqn:qasymptotic}
\end{equation}
In the above expression $C'$ is a quantity that approaches to
$C=\frac{1}{\sqrt{12\pi}}$ as $q \rightarrow 1$ and could be a function
of $J_i$ at $k^{-2}$ order.
The second term inside the cosine corresponds
to  a regularization counter term,
while the spin dependence of $C'$ may be regarded as a correction
for the measure. The integral of $\theta_1$ in the cosine is proved
to be $\sum \theta_i J_i$ \cite{SG}.

	Next we estimate \(\varrho / \sin \theta\) in the large \(k\)
and angular momenta limit.
Defining $\qc=\frac{\pi^2}{6k^2}$, then
\begin{equation}
 \b{\l J}= \left( \l J -\qc \l J^3 \right)
 \left[1+O({\l^{-2}})\right], \label{eqn:best}
\end{equation}
so that
\begin{eqnarray}
 c(J_1) &=&
 -\frac{h(\l j_1)}{16\left\{ \b{2(\l J_1+1)}
 \b{2(\l J_1-1)}\right\}^{1/2}
 F_q(\l J_1, \l J_2, \l J_3)F_q(\l J_1, \l J_5, \l J_6)}
 \nonumber \\
 &=&-\frac{\l^5 \left( h_c+\qc h_q \right) \bigl\{
 1+\qc\left( 8J_1^2+2\left( J_2^2+J_3^2+J_5^2+J_6^2 \right)
 \right) \bigr\}}{2\l J_1 \cdot 16 \fc{\l J_1,\l J_2,\l J_3}
 \fc{\l J_1,\l J_5,\l J_6}} \nonumber \\
 &=& \cos \theta_1 -\qc \frac{h_q+\left(
 8J_1^2+2\left( J_2^2+J_3^2+J_5^2+J_6^2 \right) \right)h_c }
 {2J_1 \cdot 16 \fc{J_1,J_2,J_3}\fc{J_1,J_5,J_6}}+O(\l^{-2})~,
\label{eqn:ccor}
\end{eqnarray}
where $h_c$ and $h_q$ are the following:
\begin{eqnarray}
 h_c&=&-2 J_1 (J_1^4 - J_1^2 J_2^2 - J_1^2 J_3^2
 + 2 J_1^2 J_4^2 - J_1^2 J_5^2  \nonumber \\
 & & \qquad \qquad -J_2^2 J_5^2 + J_3^2 J_5^2 -
 J_1^2 J_6^2 + J_2^2 J_6^2 - J_3^2 J_6^2) \\
 h_q& =&4 J_1 (4 J_1^6 - 3 J_1^4 J_2^2 - J_1^2 J_2^4 - 3 J_1^4
 J_3^2 -J_1^2 J_3^4 + 12 J_1^4 J_4^2 + 2 J_1^2 J_4^4 - 3 J_1^4 J_5^2
 \nonumber \\
 & & \qquad - 8 J_1^2 J_2^2 J_5^2 - J_2^4 J_5^2 + 2 J_1^2 J_3^2 J_5^2
 +  J_3^4 J_5^2 - J_1^2 J_5^4 - J_2^2 J_5^4 + J_3^2 J_5^4
 -3 J_1^4 J_6^2 \nonumber \\
 & & \qquad \  + 2 J_1^2 J_2^2 J_6^2 + J_2^4 J_6^2 -
  8 J_1^2 J_3^2 J_6^2 - J_3^4 J_6^2 -
 J_1^2 J_6^4 + J_2^2 J_6^4 - J_3^2 J_6^4).
\end{eqnarray}
$\fc{a,b,c}$ is the `classical' value of $\fq{a,b,c}$,
that is, the area of a triangle
whose edges are of length $a$, $b$ and $c$.

The numerator of the second term
of (\ref{eqn:ccor}) proved to be:
\begin{equation}
8J_1^3 \cdot 16 \fc{J_4,J_2,J_6} \fc{J_4,J_5,J_3} \cos \theta_4.
\end{equation}
Using the formula
\begin{equation}
 \frac{\partial V}{\partial J_1}=-
 \frac{J_1 \cdot 16 \fc{J_4,J_2,J_6} \fc{J_4,J_5,J_3} \cos \theta_4}
 {144 V},
\end{equation}
we obtain
\begin{eqnarray}
 \varrho&=&-\qc \times \frac{
 8J_1^3 \cdot 16 \fc{J_4,J_2,J_6} \fc{J_4,J_5,J_3} \cos \theta_4}
 {2J_1 \cdot 16 \fc{J_1,J_2,J_3}\fc{J_1,J_5,J_6}}\nonumber\\
&=& 24\qc \frac{\partial V}{\partial J_1} \sin \theta_1 .
\label{eqn:varrho}
\end{eqnarray}
Inserting (\ref{eqn:varrho}) into (\ref{eqn:qasymptotic}),
we conclude that
\begin{equation}
I_{TV}(\mbox{positive frequency part})
 \sim \int d\mu' \exp
\left( i \frac 12 \int \sqrt{g} \left( R -\frac{8\pi^2}{k^2}
\right) \right),
\end{equation}
where $d\mu ' = d\mu + O(k^{-2})$.

	It is a plausible result that the cosmological term appears
as a regularization counter term though our approximation
for $q$-$6j$ symbol is valid only in the region $j_i \ll k$
and all the $j_i$ are uniformly large. If the sum-region with respect
to edge-lengths is unbounded, any configuration with high deficit angles
would be allowed. In our case the summation is truncated at length $k$
so that the configuration is chopped off if it is spiky enough;
consequently, the virtual processes with large volume are suppressed.

	A crucial question is the following:
``what does it mean if you gather only negative frequency factors
to obtain a `Lorentzian' action
($i\times$classical action), though you have been considering
a Euclidean system?''
In $2+1$-dimension the Einstein action coincides with the Chern-Simons
(CS) action with gauge group $G$ under some
appropriate field-identification,
where $G$ is the isometry group of space-time, {\em i.e.}
$G=\mbox{\rm ISO}(2,1)$, $\mbox{\rm SO}(3,1)$ and $\mbox{\rm SO}(2,2)$
if space-time is Minkowski, de-Sitter and anti-de-Sitter,
respectively \cite{Wi2+1}. However, in 3-dimension with Euclidean
signature the Lagrangian for the $\mbox{\rm SO}(4)$ CS theory,
which is supposed to be the Euclidean version
for space-time with positive cosmological constant,
is pure imaginary; hence the equivalence to the Einstein gravity is
subtle \cite{Wi2+1}. On the other hand, we started from
the TV invariant and interpreted it as the path-integral
in the semi-classical continuum limit.
Since amplitude for any process is topologically invariant,
so should be the action. Therefore it would be better
to consider the Regge action appeared in our model
as the CS action, rather than the Einstein action.
The strange $i$ factor may indicate the subtlety
in the correspondence between the Euclidean gravity
and a CS theory.

       Indeed, there is an evidence that the TV invariant
is related to a CS theory. If one evaluates $I_{TV}$
in a 3-manifold
$F\times{[0,1]}$ where $F$ is a triangulated 2-surface, then
$I_{TV}$ naturally
induces a representation of the modular group of $F$ \cite{TV}.
Comparing the representation with that of the Jones polynomial,
Turaev and Viro conjectured that their invariant would be related
to the `square' of the Hilbert space of a topological field theory
for the Jones polynomial, {\em i.e.} the $\mbox{\rm SU}(2)$
CS theory with Wilson lines \cite{WiJones}.
Recently, it has been shown in ref.\cite{OS}
that the TV invariants can be constructed
from the $\mbox{\rm SO}(4)$ ($=\mbox{\rm SU}(2)\times\mbox{\rm SU}(2)$)
CS theory with Wilson lines in the large $k$ limit.

	We would like to thank M. Ninomiya for very fruitful discussions
and continuous encouragement. We also thank M. Hayashi, A. Hosoya,
S. Iso, S. Sawada, J. Soda and H. Suzuki for discussions.
One of us (T. T.) also thanks A. Kirillov and H. Ooguri
for discussions, and appreciate a kind hospitality of RIMS\ 91 Project.
This work is supported
in part by JSPS and Soryuushi shogakukai.
%
%
%

\vspace{1cm}
\hspace{3cm}
\raisebox{0mm}{
\psfig{figure=tetra.ps,width=7cm,height=7cm}
}

\noindent Fig.1.
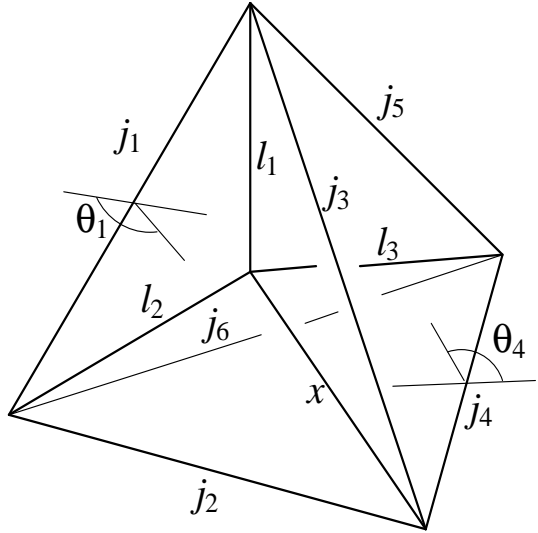
{The 6j symbol
$\left\{
\protect\begin{array}{ccc}
j_1&j_2&j_3 \\ j_4&j_5&j_6
\protect\end{array}
\right\}$ is represented by a tetrahedron whose
edges are of length $\underline{J_i = j_i + \frac 12}$.
$\theta_i$ is the angle between
the outer normal of the two faces which have the edge $J_i$ in common.
The subdivision (\ref{eq:rensubdivision}) is also illustrated.}

\message{
THE FIGURE IS AVAILABLE FROM POSTSCRIPT FILE.
SEE INSTRUCTION AT THE TOP OF THE FILE.}

\begin{thebibliography}{99}
\bibitem{PR}
G. Ponzano and T. Regge, in:{\it Spectroscopic and group
theoretical methods in physics},
 ed. F. Bloch (North-Holland, Amsterdam, 1968).
\bibitem{HP}
B. Hasslacher and M. J. Perry, {\it Phys. Lett.} {\bf B103}
(1981) 21.
\bibitem {SG}K. Schulten and R. G. Gordon,
{\it Jour. Math. Phys.} {\bf 16} (1975)
1961; {\it ibid} 1971.
\bibitem{Regge}
T. Regge, {\it Nuovo Cimento} {\bf 19} (1961) 551.
\bibitem{TV}{V. G. Turaev and O. Y. Viro, LOMI preprint.}
\bibitem {KR}
A. N. Kirillov and N. Yu. Reshetikhin, {\it in Adv. Ser. in Math.
Phys,} ed. G. Kac, vol. 7 (1988).
\bibitem{Wi2+1}
E. Witten, {\it Nucl. Phys.} {\bf B311} (1988) 46.
\bibitem{WiJones}
E. Witten, {\it Commun. Math. Phys.}{\bf 121}(1989)351.
\bibitem{OS}
H. Ooguri and N. Sasakura, KUNS-1088, August 1991.
\bibitem{Nomura1}
M. Nomura, {\it J. Math. Phys.} {\bf 30} (1989) 2397.
\end{thebibliography}
\end{document}